# An Energy-Efficient Heterogeneous Memory Architecture for Future Dark Silicon Embedded Chip-Multiprocessors

**SALMAN ONSORI, ARGHAVAN ASAD, KAAMRAN RAAHEMIFAR, AND MAHMOOD FATHY**

S. Onsori is with the Computer Engineering Department, Bilkent University, Ankara 06800, Turkey
A. Asad and M. Fathy are with the Computer Engineering Department, Iran University of Science and Technology, Tehran, Iran
K. Raahemifar is with the Electrical and Computer Engineering Department, Ryerson University, ON M5B 2K3, Canada

CORRESPONDING AUTHOR: S. ONSORI (s.onsori66@gmail.com)

**ABSTRACT** Main memories play an important role in overall energy consumption of embedded systems. Using conventional memory technologies in future designs in nanoscale era causes a drastic increase in leakage power consumption and temperature-related problems. Emerging non-volatile memory (NVM) technologies offer many desirable characteristics such as near-zero leakage power, high density and non-volatility. They can significantly mitigate the issue of memory leakage power in future embedded chip-multiprocessor (eCMP) systems. However, they suffer from challenges such as limited write endurance and high write energy consumption which restrict them for adoption in modern memory systems. In this article, we present a convex optimization model to design a 3D stacked hybrid memory architecture in order to minimize the future embedded systems energy consumption in the dark silicon era. This proposed approach satisfies endurance constraint in order to design a reliable memory system. Our convex model optimizes numbers and placement of eDRAM and STT-RAM memory banks on the memory layer to exploit the advantages of both technologies in future eCMPs. Energy consumption, the main challenge in the dark silicon era, is represented as a major target in this work and it is minimized by the detailed optimization model in order to design a dark silicon aware 3D Chip-Multiprocessor. Experimental results show that in comparison with the Baseline memory design, the proposed architecture improves the energy consumption and performance of the 3D CMP on average about 61.33 and 9 percent respectively.

## I. INTRODUCTION

Energy consumption is an essential and important constraint in embedded systems since these systems are generally restricted by battery lifetime. It is widely acknowledged that energy consumption of memory systems is a significant contributor to the overall system energy due to integration of increasingly larger memory closer to the processor [47]. Therefore, there is a critical need to considerably reduce energy consumption of memory architectures. Memory energy consists of two components: 1) leakage, and 2) energy of the read/write access. In order to reduce memory energy, both the leakage and dynamic energy should be minimized. Moreover, 42 percent of the overall energy dissipation in the 90 nm generation [1] and over 50 percent of the overall energy dissipation in 65 nm technology [4] are due to leakage. Hence, leakage energy has become comparable to dynamic energy in current generation memory modules and soon will exceed dynamic energy in magnitude if voltage and technology are furthur scaled down [3], [24].

Due to physical limitations of two dimensional integration technologies (2D IC), three dimensional chip-multiprocessors (3D CMPs) receive a lot of attention in these days [25]–[28]. 3D integration technology compare with 2D designs reduces interconnection wire length resulting in lower power consumption and shorter communication latency [23]. On the other hand, Network on Chips (NoC) architectures have been extended to the third dimension by the help of through silicon vias (TSVs) [44], [45]. 3D NoCs combine the benefits of short vertical interconnects of 3D ICs and the scalability of NoCs. Therefore, 3D NoCs have the potential to achieve better performance with higher scalability and lower power consumption.

Inorder to exploit 3D CMP and benefit from the advantages of 3D NoC, CMP architectures with 3D stacked memory system has been proposed to reduce power consumption of CMP and increase its performance [7], [35], [36], [53], [54]. Stacked traditional memory systems on the core layer may drastically degrade performance, power density and temperature-related problems [46] such as negative bias temperature instability (NBTI) [42]. For example by stacking eDRAM/DRAM on top of cores as on-chip memory, the heat generated by the core-layer can significantly aggravate the refresh power of DRAM layers. In such case, the designer needs to consider the power consumption due to the refreshing phase when designing the power management policy for stacked DRAM memory or cache. Non-volatile memories (NVMs) are newly emerging memory technology with potential application in designing new classes of memory systems due to their benefits such as higher storage density and near zero leakage power consumption [37]–[39]. Spin-transfer torque random-access memory (STT-RAM) as a promising candidate of NVM technology combines the speed of SRAM, the density of DRAM and the non-volatility of Flash memory. In addition, excellent scalability and very high integration with conventional CMOS logic are the other superior characteristics of STT-RAM [2]. Although NVMs have many benefits as described above, their drawbacks such as high write energy consumption, long latency writes and limited write endurance prevent from their direct use as a replacement for traditional memories [32], [48].

In order to overcome the aforementioned disadvantages, we use eDRAM and STT-RAM as two different types of memory banks in the stacked memory layer in a 3D eCMP. This hybrid memory architecture leads us to the best design possible exploiting the benefits of both of memory technologies. In this work, we use Non Uniform Memory Architecture (NUMA) stacked directly on top of the core layer in the proposed eCMP.

Recently, dark silicon has emerged as a trend in VLSI technology [29], [30], [49], [50]. The rise of utilization wall due to thermal and power budgets restricts active components and results in a large region of dark silicon. Uncore components, such as memory and cache subsystem, consume a significant amount of power consumption [31]. Thereby, power management of uncore components is critical for maximizing design performance in dark silicon era. We exploit 3D die-stacking and emerging NVM in this work to design high performance 3D CMP architecture for minimizing energy consumption as a solution to combat dark silicon challenge. Previous research has mainly focused on energy efficient core designs [29], [40], and the design of uncore components for reducing energy consumption has been rarely explored. Heterogeneous architectures can be a promising solution to tackle the challenges of multicore scaling in the dark silicon era because of slight improvement in CMOS technology. NVMs can be efficiently integrated with CMOS circuits in energy-efficient designs.

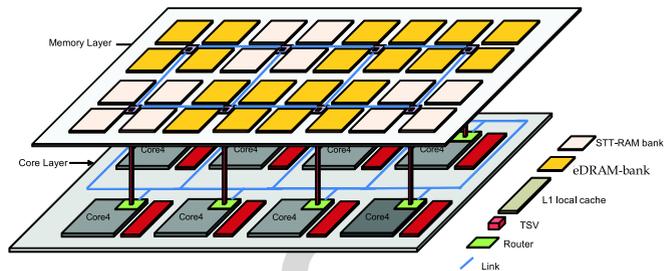

**FIGURE 1.** An overview of the proposed architecure.

To the best of our knowledge, this paper is the first work to examine an energy efficient heterogeneous memory architecture design based on a convex optimization approach for future eCMPs. We exploit 3D die-stacking and emerging NVMs to design a high performance 3D eCMP architecture to minimize energy consumption as a solution to combat dark silicon challenge for future CMP.

Figure 1 shows an overview of the proposed design using an example of an 8 homogeneous cores in the lower layer and hybrid memory architecture in the upper layer. In the proposed heterogeneous memory system, STT-RAM as a well-known candidate of NVMs is incorporated with eDRAM banks in the second layer.

This paper makes the following novel contributions:
- We provide convex optimization based platform to design a heterogeneous memory system consistsing of NVM and eDRAM memory banks.
- Our proposed model can optimally find the number of eDRAM and STT-RAM memory banks in the memory layer of the embedded 3D CMP based on the access behavior of mapped applications to minimize energy consumption.
- We demonstrate that our ILP formulation extends the lifetime of the hybrid memory architecture and provides significant energy savings in comparison with the baseline designs.
- We developed a simulator with hybrid memory and 3D NoC platform to evaluate the proposed design in embedded 3D CMP using PARSEC benchmarks.

The rest of this paper is organized as follows. Section II describes a brief background. Section III describes related work. In Section IV, the details of convex optimization-based problem and its formulation are investigated. In Section V, evaluation results are presented. Finally, the paper is concluded in Section VI.

## II. BACKGROUND
### A. STT-RAM TECHNOLOGY
STT-RAM has been one of the most popular NVM structures due to its scalability in sub-nanometer technology and the low writing current in comparison with the conventional Magnetic Random Access Memory (MRAM).

As it is illustrated in Figure 2, to perform a read operation from the STT-RAM cell, the NMOS transistor will be turned ON and a small voltage will be set between the bit line and the source line. This voltage causes a current in the magnetic tunnel junction (MTJ). The amount of this current depends on the state of the MTJ. A current sensor senses the current and compares it with a reference current. As a result, the logic value of that cell will be determined.

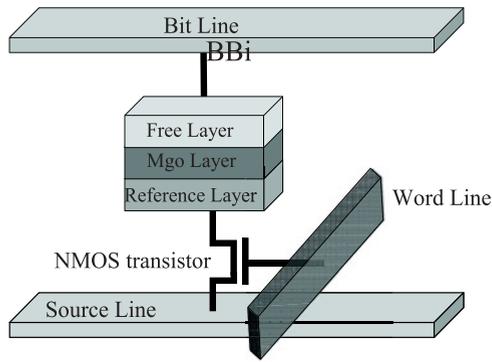

**FIGURE 2.** Structure of a STT-RAM.

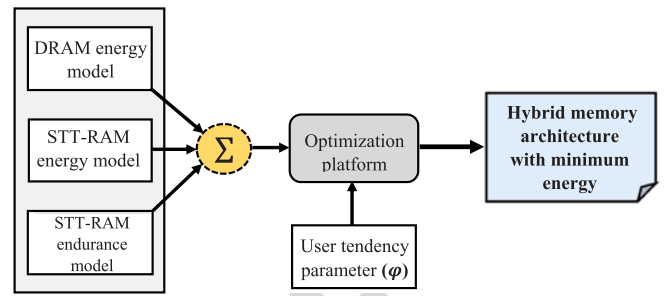

**FIGURE 3.** Overview of our model.

For a write operation, the amount of the current would vary and will depend on the cell value. In order to write a the logic value of '0' a positive current and for writing the logic value of '1', negative current is injected between bit line and source line. The amount of the current for a reliable write operation is known as threshold current which is depended on the type of material used to construct the MTJ and its shape [14], [41].

### B. 3D DIE-STACKING TECHNOLOGY
The three-dimensional integrated circuits (3D ICs) technology, where multiple silicon layers are stacked vertically, has proven to be a promising solution for increasing the number of transistors on a chip [55]. In 3D IC designs, the critical paths can be significantly shortened and the bandwidth between processor cores and memories can be greatly increased [22], [23]. In addition to the aforementioned advantages, 3D ICs also provide heterogeneous integration, on-chip interconnect length reduction, and a modular and scalable design. Thus, 3D integration is envisioned as a solution for future many-core design to tackle the memory wall problem. In this paper, we assume that the stacking approach is used for 3D embedded CMP design, in which core and memory layers are vertically stacked and connected by through silicon vias (TSVs).

### III. RELATED WORK
Numerous studies [8], [9], [33], [34] have proposed hybrid architectures, wherein the SRAM is integrated with NVMs, in order to take advantages of both technologies. Energy consumption is still a primary concern in embedded systems since they are limited by battery constraint. Several techniques have been proposed to reduce energy consumption of hybrid memory architectures in embedded systems. Fu *et al.* [12] presented a technique to improve energy efficiency through a sleep-aware variable partitioning algorithm for reducing the high leakage power of hybrid memories.

Hajimiri *et al.* [11] proposed a system-level design approach that minimizes dynamic energy of a NVM-based memory through content aware encoding for embedded systems. Our work is different from all the prior works as we focus on placement of eDRAM and STT-RAM banks in a stacked memory architecture in future CMPs to minimize energy consumption using a convex optimization based approach.

As mentioned before, there are some obstacles for employing STT-RAM without integration with tradi-tional technologies in modern memory systems. One of these obstacles is the limited number of write operations. After number of write operations has reached its limit, it is not possible to write another value into a STTRAM cell, and only the stored values can be read [43]. A number of researches presented different techniques to address the endurance problem of NVMs. Qureshi *et al.* [10] proposed wear leveling techniques for a PRAM-based memory system to enhance the lifetime. Wang *et al.* [5] proposed an algorithm to evenly distribute write events in the address space of scratchpad memory to extend the endurance of NVM. Luo *et al.* [6] presented a writing technique called Min-Shift to reduce the total number of writes to NVM and to enhance the lifetime of NVM. Hu *et al.* [13] proposed a software wear leveling technique to extend the lifetime of NVM in hybrid memory structure of embedded systems. However, our paper is the first work to propose an endurance model for NVM technology. This endurance model is used as a constraint in the proposed optimiza-tion problem to design a high endurance heterogeneous memory system with minimum energy consumption.

### IV. OPTIMIZATION MODEL
In this section, we formulate our energy optimization problem to design a minimum energy heterogeneous memory structure in embedded 3D CMP. Figure 3 shows block diagram of our model for designing the proposed hybrid memory with minimum energy consumption.

The outputs of our optimization problem are 1) finding the optimal number of eDRAM and STT-RAM memory banks based on the memory access behavior of mapped applications with respect to the endurance constraint, 2) the appropriate placement of eDRAM incorporated with STT-RAM banks in the memory layer to minimize energy consumption.

DRC and STC represent our optimization variables. These two binary variables indicate that a particular memory bank in the proposed design is either an eDRAM or a STT-RAM bank. Our convex optimization model finds DRC and STC variables for each banks in the second layer. Based on these variables, the hybrid memory layer is constructed (Figure 4). After constructing the second layer and knowing actual placement of eDRAM and STT-RAM banks on it, we can count the number of banks and hence we can find the optimal number of each memory technology in our design.

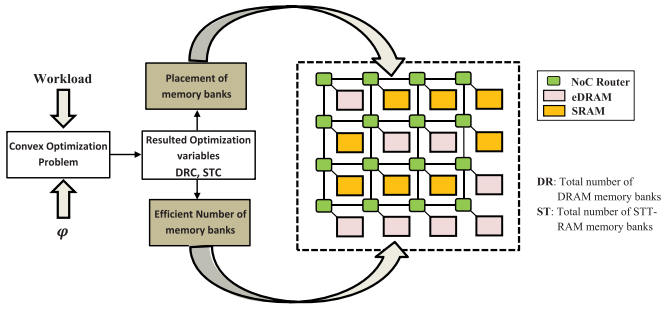

**FIGURE 4.** Construction of hybrid memory layer based on optimization variables.

Table 1 gives the constant terms used in our convex formulation. To solve the models, we used CVX [15], an efficient convex optimization solver.

Assuming that $P$ denotes the total number of processor cores, $DR$ the total available number of eDRAM memory banks, $ST$ the total available number of STT-RAM memory banks, $(C_X, C_Y)$ the dimensions of the chip, $(P_X, P_Y)$ the dimensions of the processor core. In this work, $DR$ and $ST$ are equal to $P$; however, these numbers can be different values. Our approach uses 0–1 variables to specify the coordinates of each memory bank and processor core. Note that, we do not consider application mapping in our proposed model and applications are randomly mapped to cores in the core layer.

We use $DRC$ and $STC$ to identify the coordinates of a memory bank. We have two types of memory banks, eDRAM and STT-RAM, so we have two variables.

- $DRC_{dr,x,y,l}$: indicates whether an eDRAM bank is in $(x, y)$ in layer $l = 2$.
- $STC_{st,x,y,l}$: indicates whether a STT-RAM bank is in $(x, y)$ in layer $l = 2$.

The mapping between coordinates and blocks in the second layer are ensured by variables *DRMap* and *STMap* for the eDRAM and STT-RAM memory banks, respectively. That is,

- $DRMAP_{dr,x,y,l}$: indicates whether coordinate $(x, y)$ is assigned to an eDRAM bank in layer $l = 2$.
- $STMAP_{st,x,y,l}$: indicates whether coordinate $(x, y)$ is assigned to a STT-RAM bank in layer $l = 2$.

A memory bank needs to be assigned to a unique coordinate. In Equation (1), $i$ and $j$ correspond to the $x$ and $y$ coordinates, respectively.

$$\sum_{i=1}^{C_X-1} \sum_{j=1}^{C_Y-1} (DRC_{dr,i,j,l} + STC_{st,i,j,l}) <= 1, \quad \forall dr, \forall st, \ l=2, \quad (1)$$

**TABLE 1.** Constant Terms Used in Our Optimization Problem. The Values of $FREQ_{p,m,r}$ and $FREQ_{p,m,w}$ Are Obtained by Collecting Statistics Through Simulation the Code and Capturing Accesse to Each Storage block.

| Constant | Definition |
|---|---|
| $P$ | Number of cores in the core layer |
| $DR$ | Total number of eDRAM memory banks |
| $ST$ | Total number of STT-RAM memory banks |
| $C_X, C_Y$ | Dimensions of the chip |
| $P_X, P_Y$ | Dimensions of a core |
| $R_X, R_Y$ | Dimensions of an eDRAM memory bank |
| $T_X, T_Y$ | Dimensions of a STT-RAM memory bank |
| $N$ | The number of lines in STT-RAM memory bank |
| $l$ | Index of layers in the 3D CMP |
| $FREQ_{p,m,r}$ | Number of read access to memory bank m by core p |
| $FREQ_{p,m,w}$ | Number of write access to memory bank m by core p |
| $E_{read_{dr}}, E_{write_{dr}}$ | Dynamic energy consumption per read and write access by the eDRAM memory bank |
| $E_{read_{st}}, E_{write_{st}}$ | Dynamic energy consumption per read and write access by the STT-RAM memory bank |
| $\varphi$ | Using STT-RAM versus eDRAM ratio |
| $\tau_{dr}^r, \tau_{dr}^w$ | Read and write latency of eDRAM bank |
| $\tau_{st}^r, \tau_{st}^w$ | Read and write latency of STT-RAM cache bank |
| $P_{static_{dr}}$ | Static power consumed by each eDRAM memory bank at maximum temperature limit |
| $P_{static_{st}}$ | Static power consumed by each STT-RAM memory bank at maximum temperature limit |
| $STTLine_{endurance}$ | Maximum write number for each line of STT-RAM memory bank |

$$STMAP_{st,x,y,l} \geq STC_{st,x1,y1,l}$$
$$\forall st, x, y, x1, y1 \text{ such that} \quad (2)$$
$$x1 + T_X \geq x > x1 \text{ and } y1 + T_Y \geq y > y1, \quad l=2,$$

$$DRMAP_{dr,x,y,l} \geq DRC_{dr,x1,y1,l}$$
$$\forall dr, x, y, x1, y1 \text{ such that} \quad (3)$$
$$x1 + R_X \geq x > x1 \text{ and } y1 + R_Y \geq y > y1, \ l=2.$$

Also, the sum of used STT-RAM and eDRAM banks in the second layer is equal to $P$ as follow:

$$\sum_{x=0}^{C_X-1} \sum_{y=0}^{C_Y-1} \left( \sum_{i=1}^{DR} DTC_{i,x,y,l} + \sum_{i=1}^{ST} STC_{i,x,y,l} \right) = P, \quad l=2. \quad (4)$$

In this work, the memory banks and their associated router/controller in the upper layer are the same as size the cores in the lower layer. This will prevent VLSI problems related to layout and TSV design.

In order to prevent multiple mappings of a coordinate in our grid, we assign a coordinate in the second layer to a memory bank (eDRAM or STT-RAM).

$$\sum_{i=1}^{DR} DRMAP_{i,x,y,l} + \sum_{i=1}^{ST} STMAP_{i,x,y,l} = 1, \quad \forall x, y, \ l=2. \quad (5)$$

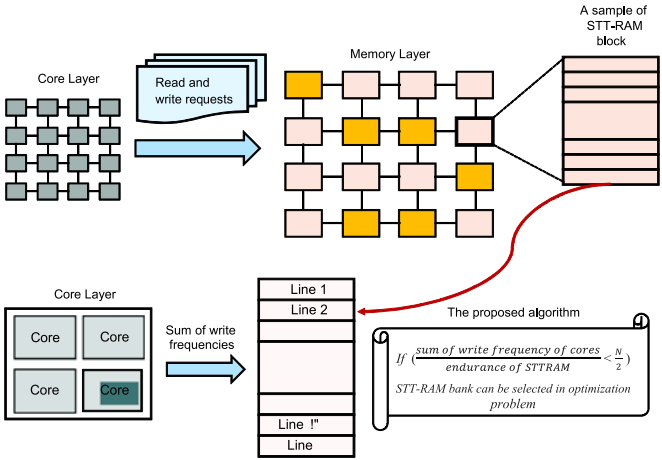

**FIGURE 5.** Overview of endurance model.

The static power dissipation depends on the temperature. Since this optimization approach is solved at design time, we consider pessimistic worst case temperature assumption and calculate $P_{static_{dr}}$ and $P_{static_{st}}$ at maximum temperature limit.

$$P_{static} = \sum_{i=0}^{C_X-1} \sum_{j=0}^{C_Y-1} \left( \sum_{k=1}^{DR} DRC_{k,i,j,l} \times P_{static_{dr}} + \sum_{k=1}^{ST} STC_{k,i,j,l} \times P_{static_{st}} \right), l = 2. \quad (6)$$

We consider endurance problem of STT-RAM in our convex model. Hence, we exploit an endurance constraint for optimal placement of eDRAM and STT-RAM memory banks. In our model, if placing a STT-RAM memory bank in a special position leads to destruction of more than half of the lines of that memory due to writing frequency of cores, STT-RAM memory bank is not chosen for that position. This endurance constraint can be expressed as follows:

$$\frac{\sum_{i=1}^{P} FREQ_{i,st,w}}{STTLine_{endurance}} \times STC_{st,x,y,2} < \frac{N}{2}, \forall x, y, st. \quad (7)$$

Figure 5 shows the overview of our endurance model.

Since STT-RAM has an endurable write threshold, we can only write a limited number of times in each line of STT-RAM. If the number of writes into one line is more than the threshold, that line will be destroyed. We assume a worst case scenario in which all write operations are written in one line until the line is destroyed and after that a new line is selected for rest of write operations. When 50% of lines in a STT-RAM memory bank have been destroyed, a new write operation only has 1/2 chance to go to a valid line which has not been already destroyed. More specifically, there is equal chance for a successful or an unsuccessful write to the STT-RAM bank. If more than half lines of a STT-RAM banks is destroyed, chance of successful write to this bank is even less than 1/2. Thus, the maximum tolerable amount to guarantee writing in a healthy line with more that 50 percent probability is N/2. Increasing this amount to a number like 3N/4, decreases our chance of writing in a healthy line of a STT-RAM bank to 1/4. On the other hand, if we decrease the amount to a number less than N/2, for example N/4, our chance to write to a healthy line will be increased to 3/4; however, it limits our design because we only can place our STT-RAM in special positions with smaller amount of write operations. We selected N/2 because it is exactly at the middle and it can make a good tradeoff for increasing endurance of STT-RAM and maintaining flexibility in our design; however, this amount can be changed based on the design's purpose.

Note that, we assume the number of lines for a STT-RAM bank is equal to N. Thus, in our endurance constraint model, if placing a STT-RAM memory bank in the special position leads to destruction of more than half lines of that memory due to writing frequency of cores, STT-RAM bank is not chosen for that position. Figure 5 illustrates the workflow of the endurance model.

Having specified the necessary constraints in our convex formulation, we next consider the objective function. The goal of our objective function is to minimize energy consumption of the stacked heterogeneous memory architecture in the target 3D CMP with respect to the endurance constraint. A weighted objective function is considered to capture its potential effects on power consumption and overall performance. This is achieved by the $\varphi$ constant which is used as a knob for choosing eDRAM versus STT-RAM bank in each $x$ and $y$ coordinates in the memory layer. As mentioned before, in comparison with eDRAM technology STT-RAM is slower and has higher density and near zero leakage power. Consequently, STT-RAM banks are applicable for memory-intensive blocks and eDRAM banks are applicable for computation-intensive blocks. Therefore, with changing $\varphi$ value, it is possible to have an optimized design based on the designer's preference. In this work, we select $\varphi = 0.5$ in the objective function. Based on this selection, STT-RAM energy function obtains half weight in comparison with the eDRAM cost function. Thus, the proposed optimization model has more freedom to choose STT-RAM banks at the memory layer. Since STT-RAM memory banks have near-zero leakage power, we can have a low power design strategy with $\varphi = 0.5$ ($\varphi < 1$ in general). The amount of $\varphi$ can be set differently for the other design purposes.

The static energy of eDRAM and STT-RAM banks for each write and read operations are defined as multiplication of their static power consumptions and read and write durations.

$$E_{static_{dr}} = \left( \tau_{dr}^r + \tau_{dr}^w \right) \times P_{static_{dr}}, \quad (8)$$

$$E_{static_{st}} = \left( \tau_{st}^r + \tau_{st}^w \right) \times P_{static_{st}}. \quad (9)$$

In Equation (10), $E_{read_{dr}}$, $E_{write_{dr}}$, $E_{read_{st}}$ and $E_{write_{st}}$ indicate dynamic energy consumed by eDRAM and STT-RAM banks per read and write access. Figure 6 shows eDRAM and STT-RAM banks in the second layer and illustrates the static and dynamic energy parameters of each memory technology. $E_{dynamic}$, the dynamic energy consumption of the proposed heterogeneous memory system is calculated as:

$$E_{dynamic} = \sum_{i=0}^{C_X-1} \sum_{j=0}^{C_Y-1} \sum_{p=1}^{P} \left( \sum_{k=1}^{DR} DRC_{k,i,j,l} \right.$$
$$\times (FREQ_{p,k,r} \times E_{read_{dr}} + FREQ_{p,k,w} \times E_{write_{dr}})$$
$$+ \sum_{k=1}^{ST} STC_{k,i,j,l} \times (FREQ_{p,k,r} \times E_{read_{st}}$$
$$\left. + FREQ_{p,k,w} \times E_{write_{st}}) \right), \ l = 2.$$
(10)

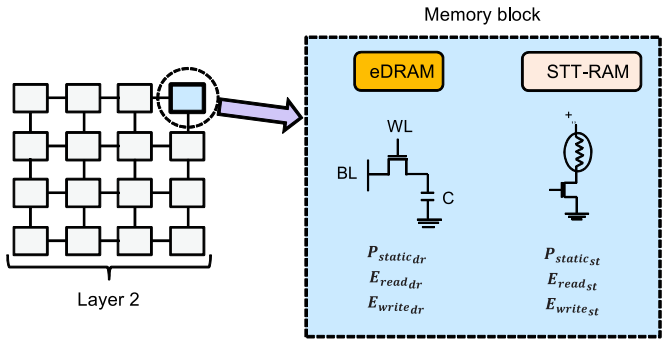

**FIGURE 6.** Energy and power parameters of a memory bank in second layer of the design.

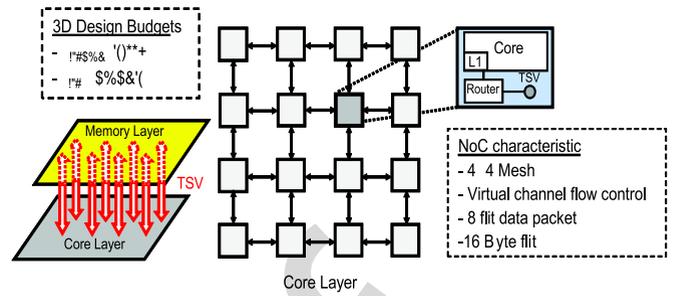

**FIGURE 7.** 3D eCMP configuration.

Consequently, our objective function can be expressed as:

$$\text{minimize } E_{Total} = (E_{static_{dr}} + E_{dynamic_{dr}}) + \varphi.(E_{static_{st}} + E_{dynamic_{st}})$$
(11)

To summarize, objective function $E_{Total}$ is minimized under constraints (1) through (10). This proposed memory system and convex optimization model is very flexible. For example in the proposed architecture, we can use other types of NVM technologies such as PCM instead of STT-RAM banks in the memory layer.

## V. EVALUATION
In this section, we first describe the experimental environment for evaluation of the proposed architecture. In the next part, different experiments are performed to quantify the advantages of the proposed architecture over the baseline architectures.

### A. EVALUATION SETUP
We used GEM5 [16] full system simulator to implement memories and cores. To simulate accurate behavior of the 3D CMP design and its NoC architecture, we integrated GEM5 with 3D-Noxim [18] which is a SystemC-based NoC simulator. We also integrated McPAT [17] with the aforementioned simulation platform in order to calculate the power consumption of the design. Furthermore, the cache capacities and energy consumption of eDRAM and STT-RAM have been estimated from CACTI [19] and NVSIM [20], respectively.

Figure 7 demonstrates the structure of the core layer and its network on chip characteristics in the proposed 3D eCMP design. Also, the simulation platform of this work is shown in Figure 8. Tables 2 and 3 list the details of system configuration for the evaluation part along with the parameters used in our experiments for eDRAM and STT-RAM memory technologies. We used multithreaded workloads in our experiments. The multithreaded applications with small working sets are selected from the PARSEC benchmark suit [21]. Moreover, $P_{budget}$ and $T_{max}$ were considered 100W and 80°C for the experimental evaluation part.

### B. EXPERIMENTAL RESULTS
In this sub-section, we evaluate the target 3D eCMP with stacked memory in four different cases: the CMP with eDRAM-only stacked memory (Baseline-eDRAM), the CMP with hybrid stacked memory that has four eDRAM banks at the middle (eDRAM-centric), the CMP with hybrid stacked memory that has same number of eDRAM and STT-RAM banks (Hybrid-symmetric), and the CMP with the proposed hybrid stacked memory based on convex optimization model. In the proposed method, we consider 16 eDRAM banks (4 MB each) and 16 STT-RAM banks (4 MB each) as the maximum available memory which can be used for designing the hybrid memory architecture. For evaluation purposes, the results of the proposed design are compared with those of the baseline designs. Baseline designs are shown in Figure 9.

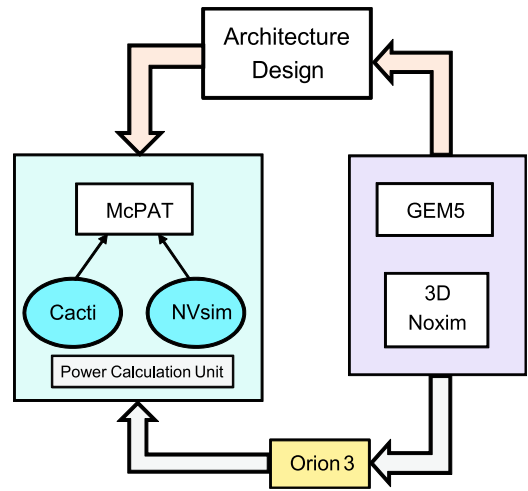

**FIGURE 8.** Simulation platform of the design.

**TABLE 2.** Specification of The Baseline eCMPs Configuration.

| Component | Description |
|---|---|
| Number of Cores | 16, 4 × 4 mesh |
| Core Configuration | Alpha21164, 3GHz, area 3.5mm$^2$, 32nm |
| Private Cache per each Core | SRAM, 4 way, 32B line, size 32KB per core |
| On-chip Memory | Baseline-eDRAM: 64MB (4MB eDRAM bank on each core) |
|  | Baseline-STTRAM: 64MB (4MB STT-RAM bank on each core) |
|  | Hybrid-symmetric: 32MB STT-RAM and 32MB eDRAM (8 STT-RAM and 8 eDRAM banks, 4MB each bank) |
|  | eDRAM-centric: 48MB STT-RAM and 16MB eDRAM (12 STT-RAM and 4 eDRAM banks, 4 MB each bank) |
|  | Hybrid proposed: the proposed hybrid memory based on the convex optimization model |
| Network Router | 2-stage wormhole switched, virtual channel flow control, 2 VCs per port, 5 flits buffer depth, 8 flits per a data packet, 1 flit per address packet, 16-byte in each flit |

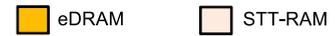

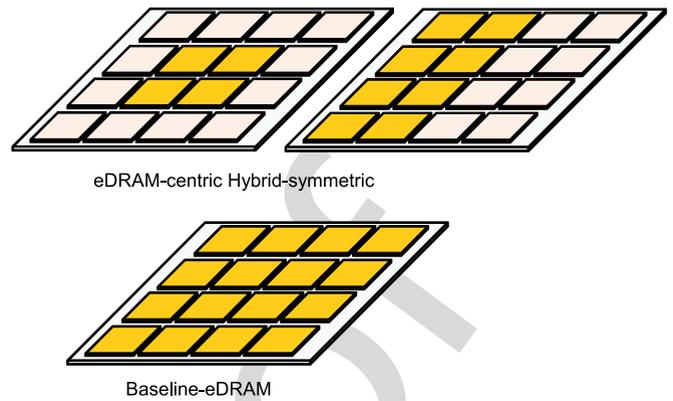

**FIGURE 9.** Different baseline designs.

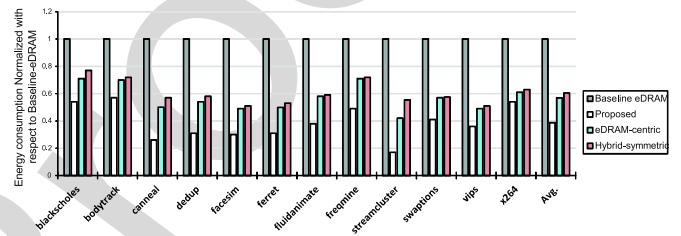

**FIGURE 10.** Comparison of energy consumption for the different baselines and the proposed memory architecture normalized with Baseline-eDRAM.

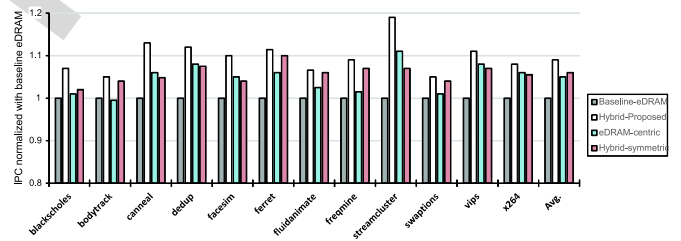

**FIGURE 11.** Comparison of instruction per cycle (IPC) for the different baselines and the proposed memory architecture normalized with Baseline-eDRAM.

Figure 10 shows the results of energy consumption for each PARSEC application. As shown in this figure, the proposed design reduces energy consumption by, on average, about 61.33, 32 and 36 percent compared with the Baseline-eDRAM, eDRAM-centric and Hybrid-symmetric designs, respectively. The educed energy consumption is due to efficient use of eDRAM and STT-RAM banks on the memory layer which is done by the proposed optimization model.

Figure 11 compares the instruction per cycle (IPC) of the proposed 3D-stacked hybrid memory architecture with the baseline designs. eDRAM and STT-RAM capacity is slightly same. Therefore, IPC differences amongst the baseline designs is due to various read and write latencies of eDRAM and STT-RAM memory technologies. Based on Table 1, although read latency in eDRAM is higher than read latency in STT-RAM, STT-RAM's write problem has a worse impact on IPC than eDRAM's read latency. For example, in Hybrid-symmetric design, half of STT-RAM banks are replaced with eDRAM banks. Hence, Hybrid-symmetric can give a higher IPC than Baseline-STTRAM design since the write problem of STT-RAM can be mitigated by eDRAM banks. Also, it is possible that Baseline-STTRAM has better IPC than Hybrid-symmetric design in read intensive benchmarks. This is because there are too many read operations in read intensive benchmarks, and this increases time required to access the memory layer due to higher read latency of eDRAM. The proposed hybrid memory architecture based on our convex optimization model has maximum IPC compared with the baseline design for all the benchmarks. Experimental results show that the proposed hybrid memory architecture gives, on average, about 9, 2.8 and 1 percent speedup on over Baseline-eDRAM, Hybrid-symmetric and eDRAM-centric designs, respectively.

**TABLE 3.** Different Memory Technology Comparisons at 65 nm.

| Technology | Area | Read Latency | Write Latency | Leakage Power at 80°C | Read Energy | Write Energy |
|---|---|---|---|---|---|---|
| 128 KB SRAM | 3.62 mm$^2$ | 2.252 ns | 2.264 ns | 131.1 mW | 0.895 nJ | 0.797 nJ |
| 512 KB STTRAM | 3.30 mm$^2$ | 2.318 ns | 11.024 ns | 16 mW | 0.858 nJ | 4.997 nJ |
| 512 KB eDRAM | 3.51 mm$^2$ | 4.053 ns | 4.015 ns | 120 mW | 0.790 nJ | 0.788 nJ |
| 2 MB PCRAM | 3.85 mm$^2$ | 4.636 ns | 23.180 ns | 31 mW | 1.732 nJ | 3.475 nJ |

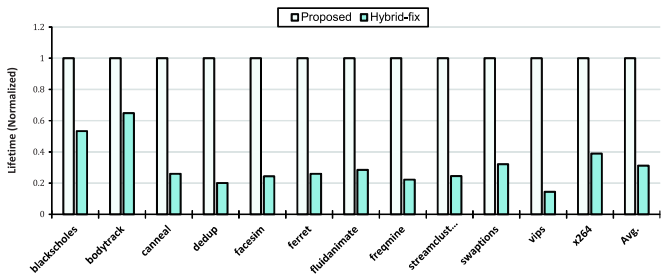

**FIGURE 12.** Expected life time comparison of the proposed design.

**TABLE 4.** Comparison of Maximum Possible Write Number for Various Memory Technologies.

| Technology | SRAM | eDRAM | STT-RAM | PRAM |
|---|---|---|---|---|
| Endurance | $10^{16}$ | $10^{16}$ | $4 \times 10^{12}$ | $10^9$ |

Figure 12 compares the lifetime of the proposed design with the Hybrid-symmetric for each benchmark. We assume the endurable maximum write number for eDRAM and different NVM memory technologies are as reported in Table 4 [51], [52].

To evaluate the lifetime, we assume that each benchmark continuously runs until one of the memory lines in each memory bank exceeds the number of maximum endurable writes (shown in Table 4). Figure 12 shows that the lifetime of our proposed heterogeneous memory architecture is higher than the lifetime of the baseline designs for all the benchmarks. The proposed hybrid memory architecture yields on average 3.03 times (and up to five times) improvement in lifetime when compared with Hybrid symmetric memory design. Thus, our hybrid memory architecture results in a more reliable 3D eCMP design due to opthe timal number and optimal placement of STT-RAM and eDRAM banks on the memory layer.

Figure 13 shows the results of energy delay product (EDP) for each PARSEC application. As shown in this figure, based on the energy consumption and performance improvement of the proposed architecture, our design improves the EDP by about 65 percent on average compare with the baseline-eDRAM.

The generated hybrid memory architectures for the canneal and fluidanimate benchmarks based on the proposed convex optimization model are shown in Figure 14. As we mentioned earlier, the number and placement of banks for each memory technology (eDRAM and STT-RAM) in the memory layer are calculated in order to minimize the performance cost function of the 3D eCMP while keeping the power budget at the satisfactory level. In other words, it depends on distribution of threads/applications on the core layer for each individual benchmark based on the convex optimization model.

## VI. CONCLUSION

In this work, we proposed a convex optimization based model to design a heterogeneous memory organization using eDRAM and STT-RAM memory banks in order to minimize energy consumption of future 3D eCMPs. We proposed an endurance model for NVM memory technologies in our optimization problem to design a reliable hybrid memory structure for the first time. The experimental results showed that the proposed method improves energy-delay product by 65 percent on average when compared with the traditional memory designs in which single technology is used. Furthermore, our 3D eCMP yields on average 9 percent performance improvement when compared with baseline designs.

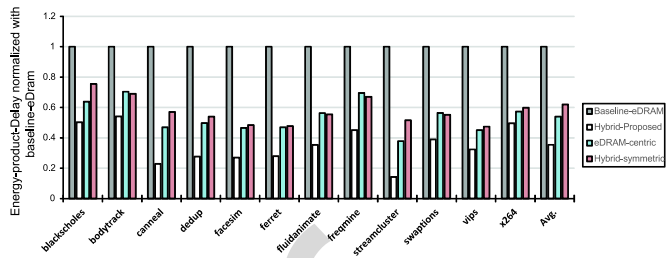

**FIGURE 13.** Comparison of energy×delay consumption for the different baselines and the proposed memory architecture normalized with Baseline eDram.

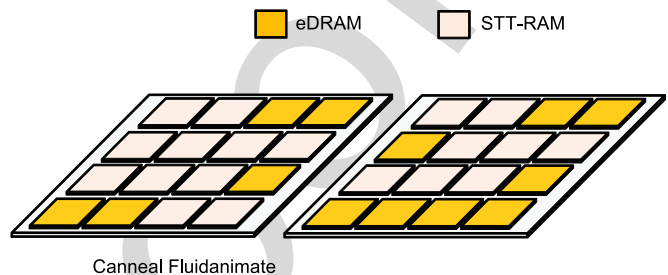

**FIGURE 14.** Hybrid memory layer for the canneal and fluidanimate benchmarks based on the proposed convex optimization model.

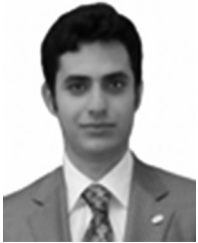

**Salman Onsori** received the BS degree in computer engineering (hardware) from the Shahed University, Iran, in 2010 and the MS degree in computer architecture from the Shahid Beheshti University, Iran, in 2013. He is currently working toward the PhD degree in the Bilkent university, Turkey. His current research interests include design of the emerging non-volatile memory and cache architectures, 3D chip-multi processors and embedded systems as well as their hardware modelling.

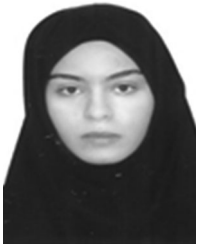

**Arghavan Asad** received the MS degree in computer architecture from the Iran University of Science and Technology, Tehran, Iran, in 2012. She is currently working toward the PhD degree at the Iran University of Science and Technology. Her research interest include interconnection network, low power hardware and memory hierarchy design.

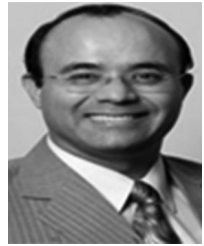

**Kaamran Raahemifar** received the BSc degree from the Sharif University of Technology, the MS degree from the Waterloo University, and the PhD degree from the Windsor University, all in electrical and computer engineering. He is currently a professor in the Department of Computer Engineering at Ryerson University. His research interests are in the areas of optimization in engineering, modeling, simulation, design and VLSI circuits.

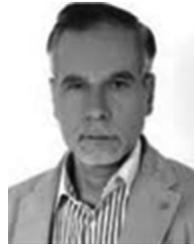

**Mahmood Fathy** received the BS degree in electronics from the Iran University of Science and Technology, Tehran, Iran, in 1985, the MS degree in computer architecture from the Bradford University, West Yorkshire, United Kingdom, in 1987, and the PhD degree in image processing and compute architecture from the University of Manchester Institute of Science and Technology, Manchester, Unnited Kingdom, in 1991. Since 1991, he has been an associate professor with the Department of Computer Engineering, Iran University of Science and Technology. His research interests include the quality of service in computer networks.